\documentclass[12pt]{iopart}
\usepackage{graphicx}
\usepackage{bm}

\begin{document}

\title{Electronic transport through a graphene-based
ferromagnetic/normal/ferromagnetic junction}
\author{Jiang-chai Chen$^1$, Shu-guang Cheng$^2$, Shun-Qing Shen$^{3}$, and
Qing-feng Sun$^{1,*}$}

\address{
$^1$Beijing National Lab for Condensed Matter Physics and Institute
of Physics, Chinese Academy of Sciences, Beijing 100190, China\\
$^2$Department of Physics, Northwest University, Xi'an 710069,
China \\
$^3$Department of Physics, and Center of Theoretical and
Computational Physics, The University of Hong Kong, Hong Kong,
China}

\ead{sunqf@aphy.iphy.ac.cn}

\begin{abstract}
Electronic transport in a graphene-based
ferromagnetic/normal/ferromagnetic junction is investigated by means
of Landauer-B\"{u}ttiker formulism and the nonequilibrium Green's
function technique. For the zigzag edge case, the results show that
the conductance is always larger than $e^{2}/h$ for the parallel
configuration of lead magnetizations, but for the antiparallel
configuration the conductance becomes zero because of the
band-selective rule. So a magnetoresistance (MR) plateau emerges
with the value 100\% when the Fermi energy is located around the
Dirac point. Besides, choosing narrower graphene ribbons can obtain
the wider 100\% MR plateaus and the length change of the central
graphene region does not affect the 100\% MR plateaus. Although the
disorder will reduce the MR plateau, the plateau value can be still
kept about 50\% even in a large disorder strength case. In addition,
when the magnetizations of the left and right leads have a relative
angle, the conductance changes as a cosine function of the angle.
What is more, for the armchair edge case, the MR is usually small.
So, it is more favorable to fabricate the graphene-based spin valve
device by using the zigzag edge graphene ribbon.
\end{abstract}


\maketitle

\section{Introduction}

Graphene, a two-dimensional single-layer crystal of carbon atoms
arraying in a honeycomb lattice, is a novel and exciting material
\cite {nmat6183}. The low energy excitation of a graphene has a
linear dispersion relation, and the dynamics of the charge carriers
obeys the massless Dirac-like equation \cite{prl532449}. Because of
the unique band structures, the graphene has many peculiar
properties, such as that the Hall plateaus occur at half-integer
multiples of $ge^{2}/h$ with the spin and valley degeneracy $g=4$,
and the conductivity at the zero magnetic field has a non-zero
minimal value \cite{nat438197}. For the neutral graphene, its Fermi
level is located at the Dirac points, the corners of the hexagonal
first Brillouin zone. In experiment, a gate voltage can be used to
tune the Fermi level, which can be above or below the Dirac points
\cite {sci317638}. In addition, graphene also exhibits many
excellent transport characteristics: the high mobility
\cite{sci306666}, long spin relaxation length \cite{nat448571}, and
stable behaviors under ambient conditions. At room temperature, its
mobilities can be above $\mathrm{10^{4}cm^{2}/V\cdot s}$, implying
that the mean free path can be as long as a few hundred nanometers.
Because of weak spin-orbit coupling \cite{prl95226801} and low
hyperfine interaction \cite{nanolett81011}, its spin relaxation
length can reach the order of the micrometer at room temperature.
Thus, graphene may be an excellent candidate for microelectronic
applications, in particular for spintronic applications
\cite{sci2821660}.

In general, the charge carriers in graphene are not spin-polarized.
For spintronic applications, people try to inject the spin-polarized
current or to induce the spin-polarized carriers in the graphene.
Recently, many works are focusing on this issue. For example, Haugen
\emph{et al} \cite{prb77115406} suggested that the spin-polarized
carriers can be realized by growing the graphene on a ferromagnetic
(FM) insulator (e.g. EuO). Owing to the magnetic proximity effect,
an exchange split between the spin-up and spin-down carriers in the
graphene is induced, and then the carriers are spin-polarized. Based
on the first-principles calculations, Son \emph{et al}
\cite{nat444347} predicted that the zigzag graphene nanoribbon
becomes a half metal when an in-plane transverse electric field is
applied. Also, Lin \emph{et al} \cite{prb79035405} demonstrated that
electron-electron correlation in armchair graphene nano-ribbon can
generate the flat-band ferromagnetism. On the experiment side, a
large spin injection into the graphene has been realized by
connecting the graphene to an FM electrode
\cite{nat448571,itm422694,prb79081402}. Furthermore, several groups
\cite{prb79081402,PRL102137205} have performed nonlocal
magnetoresistance(MR) measurements, in which a net spin current is
brought between the injector and detector.

The well-known application in spintronics is the spin-valve effect
\cite{rmp76323}, in which the resistance of devices can be changed
by manipulating the relative orientation of the magnetizations.
Motivated by the spintronic application with the novel material, the
spin-polarized transport through graphene are currently attracting a
great deal of attentions \cite{arxiv09030407, prl102136810,
pla372725, prb76205435, prb79045405, natn3408, JPCM1}. Using the
tight-binding model, Brey and Fertig \cite{prb76205435} studied the
MR of the FM/graphene/FM junction in the limit of infinite width.
They found that the MR is rather small since the conductivity is
weakly dependent on the relative magnetization orientations of FM
leads. Ding \emph{et al} \cite{prb79045405} investigated the similar
device with two FM leads by a continuous model. The results showed
that the MR versus the bias exhibits a cusp around the zero bias in
absence of external magnetic field and oscillating behavior at the
high magnetic field. Based on the first-principles calculation, Kim
and Kim \cite{natn3408} predicted that a graphene-based spin-valve
device could have a high MR.

\begin{figure}[tbp]
\centering
\includegraphics[bb=120 241 327 533,width=8.5cm]{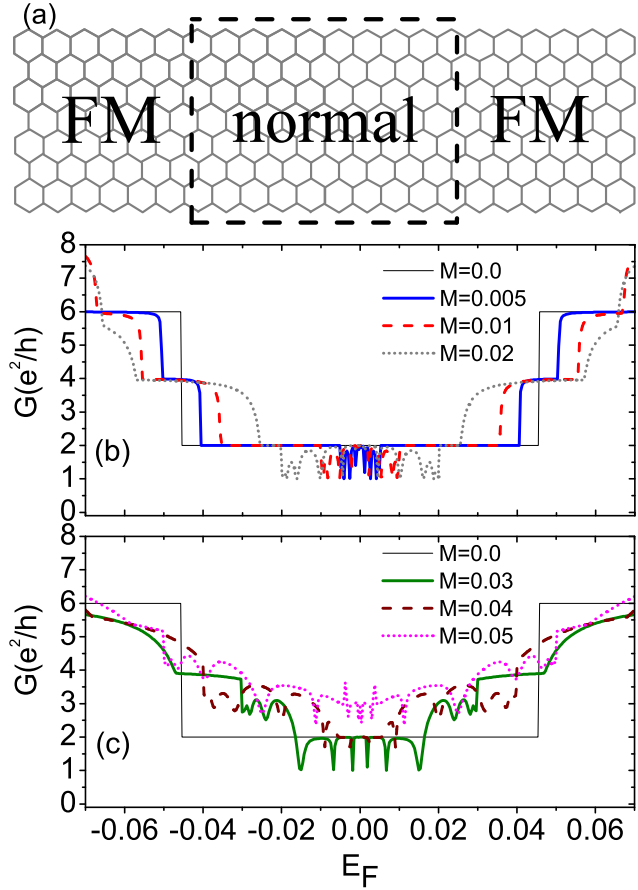}\\
\caption{(Color online)\textbf{(a)}.The schematic of a
graphene-based FM/normal/FM junction. \textbf{(b)} and \textbf{(c)}:
the conductance $G$ vs. the Fermi energy $E_{F}$ for the different
magnetization $M$ at the parallel configuration ($\protect\theta
=0$). The size of the central graphene is $N=50$ and $L=10$. }
\label{fig1}
\end{figure}

In this paper, we study the conductance and MR of a graphene-based
spin valve. The device consists of a graphene nanoribbon coupling to
two FM leads, as shown in figure~\ref{fig1}(a). Here the width of
the device is finite in the order of 10 nanometers, i.e. the size
effect is considered. In recent experiments \cite{sci3191229}, a few
10-nanometer or sub-10-nanometer graphene nanoribbons have already
been fabricated successfully. For a finite-width graphene
nanoribbon, the wave vectors along the confined direction are
discrete, and the transverse subbands emerge. The characteristics of
the subbands are strongly dependent  on the chirality of the
graphene-nanoribbon edge \cite{nanores1361}, e.g. the zigzag edge or
armchair edge. So the conductance and MR should be strongly
dependent on the boundary condition of graphene nanoribbon. In
addition, we consider that two FM leads also have the hexagonal
lattice structure as the graphene. In other words, the leads are
graphene-based FM or called the FM graphene. On the experiment, the
FM electrode (e.g. the cobalt electrode) usually overlaps on the
graphene through a thin oxide layer \cite{nat448571,prb79081402}.
Because of the magnetic proximity effect and the Zeeman effect, an
exchange split in the graphene underneath the oxide layer is
induced, then its carriers are spin-polarized. So the graphene
covered with the FM electrode has the magnetization. Then the
spin-polarized charge carriers, driven by a bias, travel from one FM
graphene through the central normal graphene to another FM graphene.

In the tight-binding model the Landauer-B\"{u}ttiker formula and
non-equilibrium Green's function method are applied to calculate the
conductance and MR. For the zigzag graphene ribbon case, we found
that when the graphene nanoribbon is narrow enough and its
separation $\triangle $ between the first subband and the zeroth
subband is larger than the exchange split energy $M$, the
conductance for the antiparallel magnetization configuration can
almost be zero in a quite large energy window around the Dirac
point. But for the parallel configuration the conductance is always
larger than $e^{2}/h$, so the MR exhibits a plateau with the value
$100\%$. As the width of the nanoribbon increases, the sub-band
separation $\triangle $ gradually decreases, and the MR can keep at
the value $100\%$ at the beginning of $\triangle>M$, then decreases
when $\triangle <M$. In the presence of the disorder, the MR
slightly drops, but even in quite large disorder it can keep the
value $50\%$, which is still much larger than 10\%, the lower limit
of the giant magnetoresistance effect. What is more, for the
armchair graphene nanoribbon case, the MR is always small regardless
of the width of nanoribbon. By comparing two cases of graphene
ribbons with different edges, it is more reasonable to apply zigzag
graphene ribbons to the spin-valve devices.

The rest of the paper is organized as follows: In Section~\ref{II},
we describe the model and the details of the calculations. In Sections \ref%
{III} and \ref{IV} we will give the numerical results of the conductance and
MR for the zigzag and armchair edge cases, respectively. Finally, a brief
conclusion is presented in Section \ref{V}.

\section{Model and formulations}

\label{II}

We consider a graphene-based spin valve (as shown in
figure~\ref{fig1}(a)) which consists of a central normal graphene
strip and two FM graphene ribbons. The total Hamiltonian $H$ of the
device can be divided into four terms, $H=H_{C}+H_{L}+H_{R}+H_{T}$,
where $H_{C}$ describes the central normal graphene region, $H_{L}$
and $H_{R}$ are the Hamiltonian of the left and right FM-graphene
leads, respectively, and $H_{T}=H_{TL}+H_{TR}$ is the coupling
Hamiltonian of central region to the left and right leads. In the
tight-binding approaximation\cite{saito}, the Hamiltonians $H_{C}$,
$H_{L}$, $H_{R}$, and $H_{T}$ can be written as
\begin{eqnarray*}
H_{C} &=&\sum_{i\in C}\epsilon _{C}a_{i}^{\dagger } \sigma _{I}
a_{i}-\sum_{<ij>(i,j\in C)}{(ta_{i}^{\dagger } \sigma _{I} a_{j}+\emph{h.c.})%
}, \\
H_{\alpha =L,R} &=&\sum_{i\in \alpha }a_{i}^{\dagger }(\epsilon _{\alpha
}\sigma _{I}+\boldsymbol{\sigma }\cdot \mathbf{M}_{\alpha })a_{i} \\
&&-\sum_{<ij>(i,j\in \alpha )}{(ta_{i}^{\dagger }\sigma _{I}a_{j}+\emph{h.c.}%
)}, \\
H_{T} &=&-\sum_{<ij>(i\in C,j\in L,R)}{(ta_{i}^{\dagger }\sigma _{I}a_{j}+%
\emph{h.c.})},
\end{eqnarray*}
where $s=\uparrow ,\downarrow $ represents the spin of electrons; $%
a_{is}^{\dagger }~(a_{is})$ creates (annihilates) an electron with
spin $s$ on site $i$, and $a_{i}={a_{i\uparrow} \choose
a_{i\downarrow}}$; $<$$ij$$>$ stands for a nearest-neighbor pair. $
\boldsymbol{\sigma }=(\sigma _{x},\sigma _{y},\sigma _{z})$ are the
Pauli matrices and $\sigma _{I}$ is a $2\times 2$ unit matrix.
$\epsilon _{C}$, $\epsilon _{L}$, and $\epsilon _{R}$ are the
on-site energies (i.e. the energy of the Dirac point) in the center
region, left and right FM leads, respectively, which can be tuned by
the gate voltage. The size of the center graphene region is
described by the width $N$ and length $L$. In figure~\ref {fig1}(a),
a zigzag edge graphene nanoribbon with $N=4$ and $L=9$ is shown. The
terms including the factor $t$ in Hamiltonian describe the nearest
neighbor hopping with the hopping energy $t$. $\mathbf{M}_{L}$ and
$\mathbf{M }_{R}$ are the magnetizations of the left and right FM
leads. Here we allow that the magnetizations $\mathbf{M}_{L}$ and
$\mathbf{M}_{R}$ can be along arbitrary direction. Without loss of
generality we assume that the magnetization $\mathbf{M}_{L}$ of left
lead is along the z-axis, then $\mathbf{M} _{L}=M_{L}(0,0,1)$, and
the magnetization $\mathbf{M}_{R}$ of the right lead is along the
direction $(\theta ,\varphi )$ with $\mathbf{M}_{R}=M_{R}(\sin
\theta \cos \varphi ,\sin \theta \sin \varphi ,\cos \theta )$.

Before performing the calculation, we take a unitary transformation with $\tilde{%
a}_{i}=Ua_{i}$ for all site $i$ in the right FM lead, where the
unitary matrix $U$ is
\begin{equation*}
U=\left(
\begin{array}{ll}
\cos (\theta /2) & e^{-i\varphi }\sin (\theta /2) \\
e^{i\varphi }\sin (\theta /2) & -\cos (\theta /2)%
\end{array}%
\right) .
\end{equation*}%
Under this unitary transformation, the Hamiltonians $H_{C}$, $H_{L}$, and $%
H_{TL}$ remain unchanged, and $H_{R}$ and $H_{TR}$ vary into
\begin{eqnarray*}
H_{R} &=&\sum_{i\in R}\tilde{a}_{i}^{\dagger }(\epsilon _{R}\sigma _{I}+%
\boldsymbol{\sigma}\cdot \mathbf{M}_{R}^{\prime })\tilde{a}_{i} \\
&&-\sum_{<ij>(i,j\in R)}(t\tilde{a}_{i}^{\dagger }\sigma _{I}\tilde{a}_{j}+%
\emph{h.c.}), \\
H_{TR} &=&-\sum_{<ij>(i\in C,j\in R)}(ta_{i}^{\dagger }U\tilde{a}_{j}+\emph{%
h.c.}),
\end{eqnarray*}%
where $\mathbf{M}_{R}^{\prime }=M_{R}(0,0,1)$. After the unitary
transformation, the z-axis of the spin in the right FM lead is along
the direction of $\mathbf{M}_{R}^{\prime }$, and the Hamiltonian
$H_{R}$ is diagonal in the spin space.

The current flowing through the device can be calculated from the Landauer-B%
\"{u}ttiker formula \cite{datta}
\begin{equation*}
I=(e/h)\int d\epsilon T_{LR}(\epsilon )[f_{L}(\epsilon )-f_{R}(\epsilon )],
\end{equation*}%
where $f_{L/R}(\epsilon )=1/\{exp[(\epsilon -\mu _{L/R})/k_{B}T]+1\}$ is the
Dirac-Fermi distribution function of the left and right FM leads and $%
T_{LR}(\epsilon )=\mathbf{Tr[\Gamma _{L}G^{r}\Gamma _{R}G^{a}]}$ is the
transmission coefficient, with the line-width function $\mathbf{\Gamma
_{\alpha }(\epsilon )}=i\mathbf{[\Sigma _{\alpha }^{r}(\epsilon )-\Sigma
_{\alpha }^{a}(\epsilon )]}$ and the Green's functions $\mathbf{%
G^{r}(\epsilon )=[G^{a}(\epsilon )]^{\dagger }=1/[\epsilon
-H_{C}-\Sigma _{L}^{r}-\Sigma _{R}^{r}]}$. $\mathbf{\Sigma
_{L/R}^{r}}$ is the retarded self-energy function coupling to the
leads, which has to be calculated numerically by solving the surface
Green's function of the leads \cite{prb234997, jpf15851}. After
solving the current $I$, the linear conductance $G$ can be obtained
straightforwardly, $G=\lim_{V\rightarrow 0}dI/dV$, with the bias
$V=\mu _{L}-\mu _{R}$. At zero temperature $ G=(e^{2}/h)T(E_{F})$.
In the numerical calculation, we take the nearest-neighbor hopping
energy $t=1$ as the energy unit and adopt the Dirac point energies
$\epsilon _{L}=\epsilon _{R}=\epsilon _{C}=0$. The magnetizations
$M_{\alpha }$ in the left and right FM leads are assumed to be equal
($M_{L}=M_{R}=M$), which is usually reasonable when two FM leads are
made of the same material. The angle $\varphi=0 $ because the
conductance and MR are independent of $\varphi $.

\begin{figure}[htbp]
\centering
\includegraphics[bb=125 26 748 454,width=8.5cm,totalheight=6.5cm]{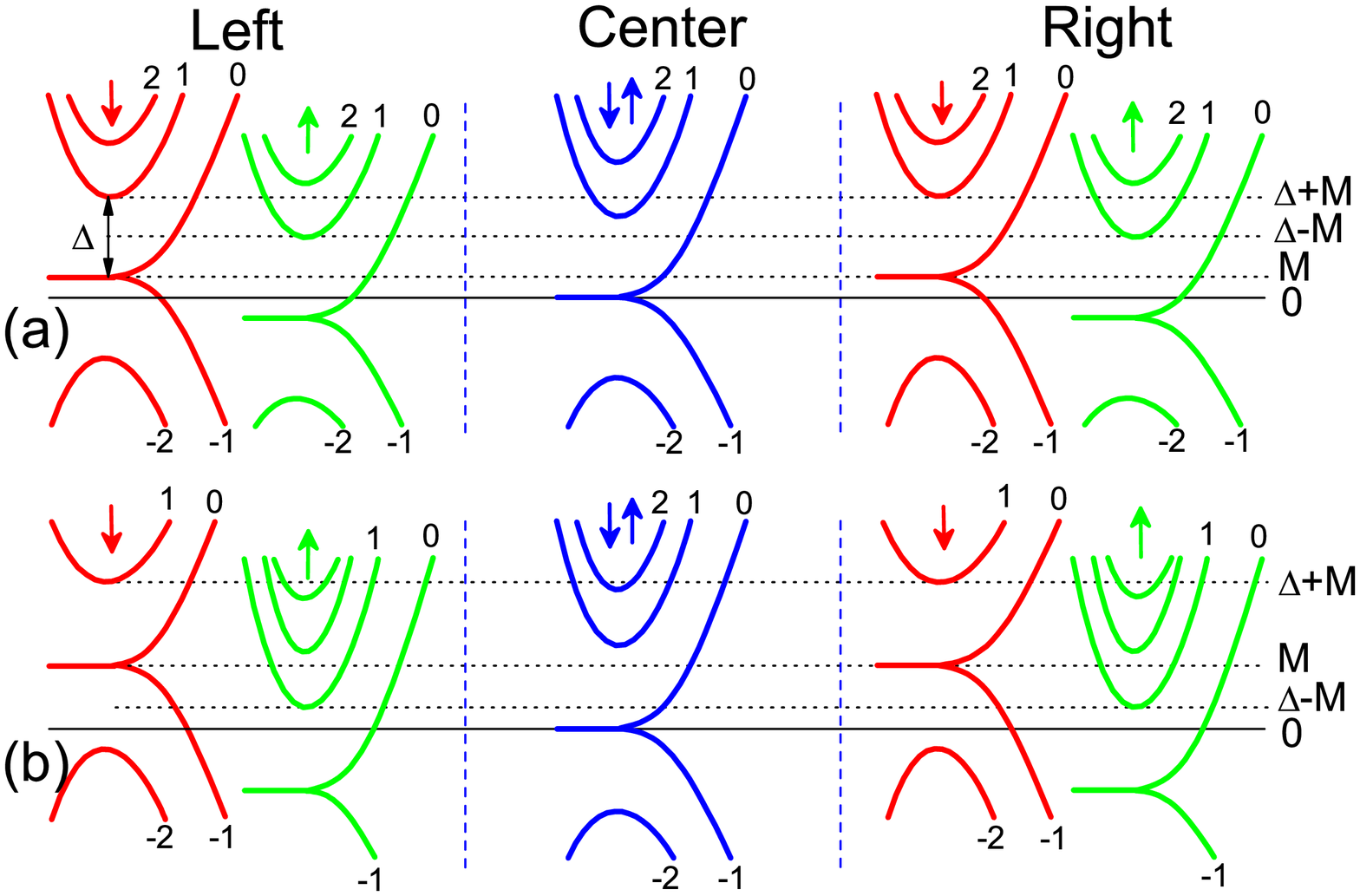}\\
\caption{(Color online) The energy band structures of the left lead,
center region, and right lead in the case of the zigzag edge at the
parallel configuration ($\protect\theta =0$). \textbf{(a)}
$0<M<\triangle /2$ and \textbf{(b)} $\triangle /2<M<\triangle $. }
\label{fig2}
\end{figure}

\begin{figure}[htbp]
\centering
\includegraphics[bb=59 24 552 516,width=8.5cm]{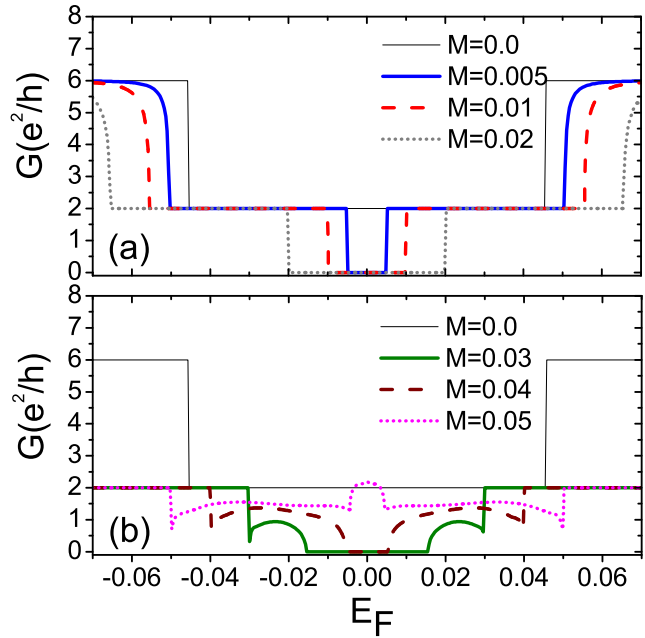}\\
\caption{(Color online) The conductance $G$ vs. the Fermi energy
$E_{F}$ for the different $M$ at the antiparallel configuration
($\protect\theta = \protect\pi $). The size of system is the same as
figure~\protect\ref{fig1}(b).}\label{fig3}
\end{figure}

\section{The case of zigzag edge}

\label{III}

In this section, we focus on the zigzag edge graphene-based spin
valves. We first investigate the conductance $G$ on the parallel and
antiparallel configurations, and then the MR and the conductance on
the arbitrary angle between magnetizations $\mathbf{M}_{L}$ and
$\mathbf{M}_{R}$.

\subsection{The conductance for the parallel configuration}

Figures~\ref{fig1}(b) and (c) show the conductance $G$ versus the Fermi energy $%
E_{F}$ for different magnetization $M$ of the parallel configuration (i.e. $%
\theta =0$). When the magnetization $M=0$, the two leads are normal
and the whole device is a flawless graphene ribbon. In this case,
the conductance $G$ exhibits the plateau structure with the plateau
values at $2,6,10,...$ (in the unit $e^{2}/h$), i.e. at the
half-integer position, $g(n+1/2)e^{2}/h$ with the degeneracy $g=4$,
due to the transverse sub-band structures. The step height of the
value $4e^{2}/h$ results from spin and valley degeneracy and the
plateaus at half-integer value originates from the fact that the
zeroth subband has the only spin degeneracy.

When the two leads are FM with a non-zero $M$, the conductance $G$
does not have the perfect plateau structure, but still obeys the
electron-hole symmetry, $G(-E_{F})=G(E_{F})$ (see
figures~\ref{fig1}(b) and (c)). In the following, we discuss the
conductance $G$ in details for $0<M<\triangle /2$ and $\triangle
/2<M<\triangle $, where $\triangle $ is the energy splitting
between the first subband and the zeroth subband (see figrue~\ref{fig2}) and $%
\triangle \approx 0.045$ for $N=50$. For $0<M<\triangle /2$, the curve of $G$
versus $E_{F}$ can be partitioned into several regions: $(0,M)$, $%
(M,\triangle -M)$, $(\triangle -M,\triangle +M)$, etc. In the first interval
of $E_{F}\in (0,M)$, the conductance $G$ oscillates with the value between $%
e^{2}/h$ and $2e^{2}/h$. In the interval of $E_{F}\in (M,\triangle
-M)$, the conductance $G$ is exactly equal to $2e^{2}/h$ and forms a
plateau. In the interval of $E_{F}\in (\triangle -M,\triangle +M)$,
the conductance approaches the value $4e^{2}/h$. With the rising of
$E_{F}$ further, the conductance becomes larger.

\begin{figure}[htp]
\centering
\includegraphics[bb=135 25 746 433,width=8.5cm,totalheight=6.7cm]{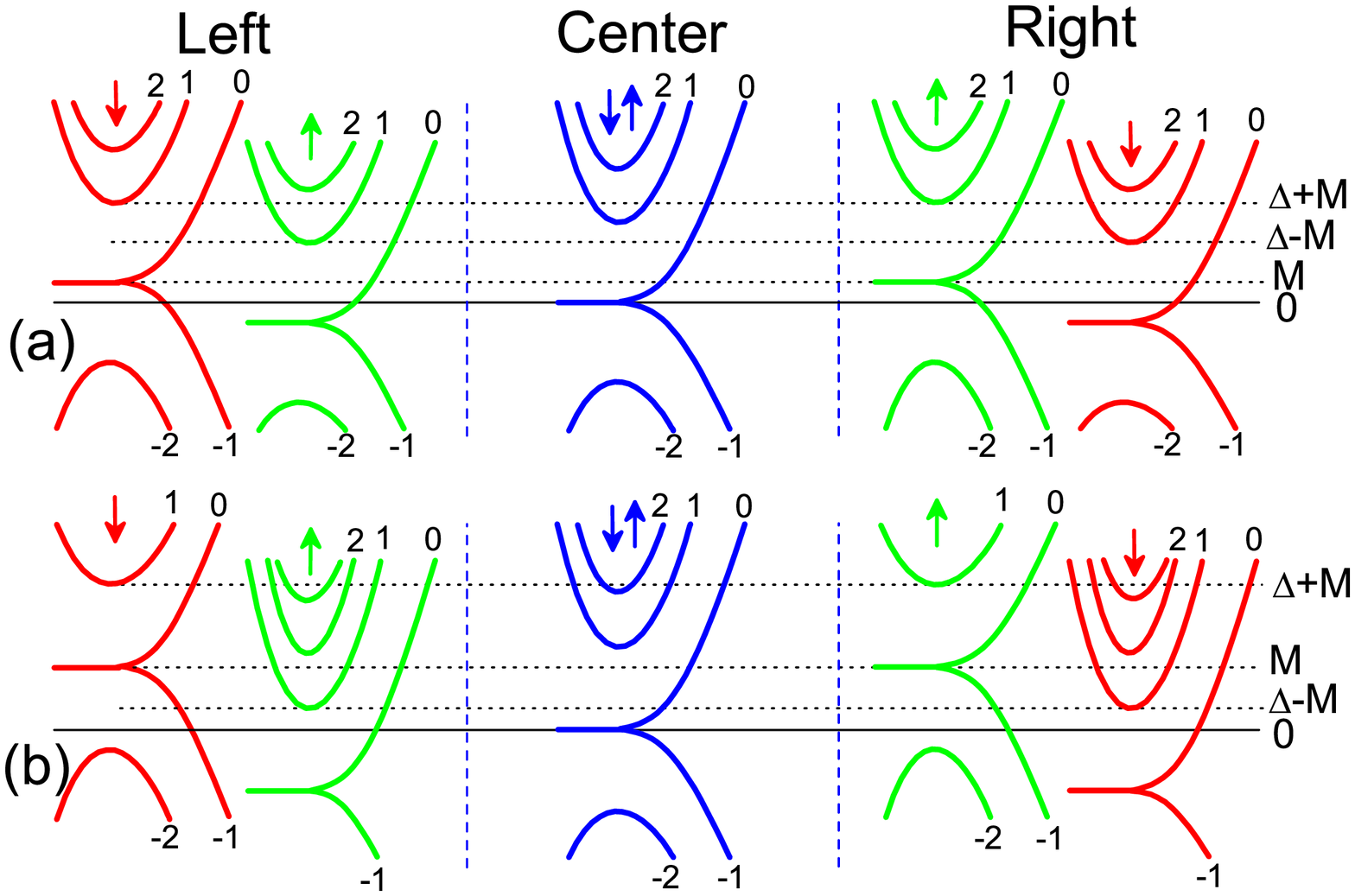}\\
\caption{(Color online) The energy band structures of the left lead,
center region, and right lead for the zigzag edge case at the
antiparallel configuration ($\protect\theta =\protect\pi $).
\textbf{(a)} $0<M<\triangle /2$ and \textbf{(b)} $\triangle
/2<M<\triangle $. } \label{fig4}
\end{figure}

In order to understand above results, we present the energy bands of
the left lead, center region, and right lead in figure~\ref{fig2},
and the band index is specified in each region. Owing to the
symmetry of bands, only the part of the moment $k>0$ is shown. The
$0$-th and $-1$-th subbands in the FM leads are nondegenerate, but
other subbands are two-fold degenerated. Because of the parity
conservation of the transverse wave function, the electrons
belonging to the even (odd) parity subbands in the left FM lead are
transported only into the even (odd) parity subbands of the right
lead, which was demonstrated in a very recent letter
\cite{prl102066803} and it is named the band-selective phenomenon.
For convenience, we use the symbol $(i,j,k)$ to denote the channel
through which electrons are transported from the $i$-th band of the
left lead, through the $j$-th band in the center region, to the
$k$-th band of the right lead, and $T_{i,j,k}$ to denote its
transmission coefficient.

With the aid of energy bands in figure~\ref{fig2}(a), we now explain
the conductance $G$ in figure~\ref{fig1}(b) with $0<M<\triangle /2$.
(i) When $E_{F}\in (0,M)$, the channel $(0,0,0)$ is open for spin-up
electrons, and $T_{0,0,0}$ is exactly 1 due to lack of the
scattering. For spin-down electrons, the channel $(-1,0,-1)$ is
open, in which the parity of the transverse wave function in the
center region mismatches with that of the left and right lead, so
the scattering exists and $0<T_{-1,0,-1}<1$. As a result, total
conductance oscillates with its value between $e^{2}/h$ and
$2e^{2}/h$. (ii) When $E_{F}\in (M,\triangle -M)$,  the channels
$(0,0,0)$ are available for both the spin-up and spin-down
electrons, and the conductance $G$ exactly equals to $2e^{2}/h$.
(iii) Furthermore, $E_{F}$ increases to the range of $(\triangle
-M,\;\triangle +M)$. Besides the channels $(0,0,0)$ of both spin
electrons, the channels $(1,j,1)$ ($j=0,1$) of the spin-up take part
in the transport, which make the conductance $G$ approximate
$4e^{2}/h$.

Now we discuss the conductance in the case of $\triangle
/2<M<\triangle $. Figure~\ref{fig1}(c) shows the conductance, and
the corresponding band structure is shown in figure~\ref{fig2}(b).
Apparently, the conductance $G$ varies more complex than that in
$0<M<\triangle /2$ (see figures~\ref{fig1}(b) and (c)). Here we are
mainly concerned about two energy regions: $E_{F}\in (0,\triangle
-M)\;\textrm{and}\;(\triangle -M,M)$. When $E_{F}\in (0,\triangle
-M)$ the transport involves the channel $(0,0,0)$ of spin-up
electrons and the channel $(-1,0,-1)$ of spin-down electrons. So the
conductance $G$ oscillates between $e^{2}/h$ and $2e^{2}/h$, which
is identical with case (i) when $0<M<\triangle /2$. For $E_{F}\in
(\triangle -M,M)$, another channel $(1,0,1)$ of spin-up electrons
joins into the transport, so the conductance is obviously enhanced
and the value lies between $ 2e^{2}/h$ and $4e^{2}/h$. Further, for
$E_{F}>M$, because more channels will be opened, the conductance
usually becomes larger than $4e^{2}/h$.

\subsection{The conductance for the antiparallel configuration}

Next, we study the antiparallel configuration with $\theta =\pi $.
The results for the conductance $G$ are shown in figure~\ref{fig3}.
One of the main characteristics is a zero conductance when $E_{F}$
is near Dirac point (i.e. $E_F=0$). This is very different from the
parallel configuration, in which $G$ is always larger than
$e^{2}/h$. Let us analyze the conductance in details in the
following. Because of $G(-E_{F})=G(E_{F})$, we only discuss the case
$E_{F}>0$. (i) For $0<M<\triangle /2$, the conductance $G$ is
exactly zero in the range $E_{F}\in (0,M)$ and becomes $2e^{2}/h$
when $E_{F}\in (M,M+\triangle )$. (ii) For $\triangle /2<M<\triangle
$, $G=0$ when $E_{F}\in (0,\triangle -M)$, while $G$ is between $0$
and $2e^{2}/h$ when $ E_{F}\in (\triangle -M,M)$ and is exactly
$2e^{2}/h$ when $E_{F}\in (M,M+\triangle )$. (iii) If $M$ increases
sequentially ($M>\triangle $), around the Dirac point a small
salient appears instead of the zero conductance range.

These characteristics of conductance can be well understood from
their band structures in figure~\ref{fig4}. From the band structures
of the antiparallel configuration the spin-up and spin-down
electrons contribute equally to the conductance, and we only discuss
the spin-up electrons in the following. (i) We discuss the case of
$0<M<\triangle /2$, and the corresponding energy band structure is
illustrated in figure~\ref{fig4}(a). When $E_{F}\in (0,M)$, only the
channel $(0,0,-1)$ is involved, in which the band is even parity in
the left lead and is odd in the right lead. Because of the
band-selective rule, its transmission coefficient $T_{0,0,-1}=0$,
and $G=0$. On the other hand, when $ E_{F}\in (M,\triangle +M)$, the
channel $(0,0,0)$ is available so that $ G=2e^{2}/h$. If further
increasing $E_{F}$, more channels are opened and the conductance is
larger. (ii) We analyze the case of $ \triangle /2<M<\triangle $,
and the corresponding energy band structure is illustrated in
figure~\ref{fig4}(b). For $E_{F}\in (0,\triangle -M)$, only the channel $%
(0,0,-1)$ is involved. So $G=0$ because of the band-selective rule. When $%
E_{F}\in (\triangle -M,M)$, the channel $(1,0,-1)$ is opened. The
parity of the transverse wave function in the center region is
different from those in the left and right leads. The incident
electrons may be scattered and $ T_{1,0,-1}$ lies between $0$ and
$1$. As a result, $0<G<2e^{2}/h$, including the spin-down electrons.
By increasing $E_{F}$ to the range $(M,\triangle +M)$, the channel
$(0,0,0)$ is opened, and the total conductance $G=2e^{2}/h$. Notice
that when $E_{F}$ is close to but less than $\triangle +M$, although
the extra channel $(2,0,0)$ is involved, the conductance $G$ still
is $2e^{2}/h$ because only the subband $0$ is available in the right
FM lead.

\begin{figure}[tbp]
\centering
\includegraphics[bb=75 30 480 413,width=8.4cm]{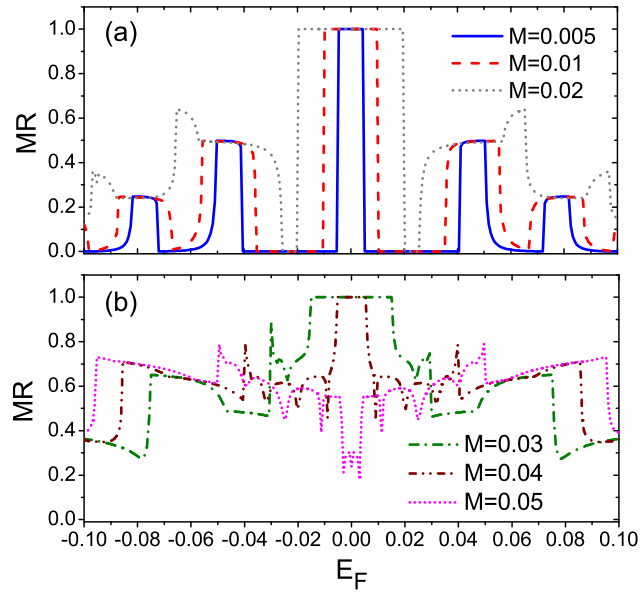}\\
\caption{(Color online) The MR as a function of $E_{F}$ for
different magnetization $M$. The size of the central region is
$N=50$ and $L=10$.} \label{fig5}
\end{figure}

\subsection{The magnetoresistance}

After obtaining the conductances $G_{P}$ and $G_{A}$ of parallel and
antiparallel configurations, the MR, defined as $%
MR=(G_{P}-G_{A})/G_{P},$ can be obtained straightforwardly.
Figure~\ref{fig5} shows the MR versus the Fermi energy $E_{F}$ for
different magnetization M. Here the MR can be very large as far as
100\% when $E_{F}$ is around the Dirac point. In particular, a
plateau with the value $100\%$ is clearly exhibited in the curve of
MR-$E_{F}$. This $100\%$ MR plateau origins from the zero
conductance in antiparallel configuration (see figure~\ref{fig3}).
It is noticed that the conductance for the parallel configuration is
still quite large ($>e^{2}/h$). This means that the present device
not only has the large MR but also has a large variance of the
conductance between the parallel and antiparallel configurations.
Thus it is a good candidate for the spin valve devices. The width of
the 100\% MR plateau is determined by magnetization $M$ and the
energy splitting $\triangle $ between the subbands. When
$0<M<\triangle /2$, its width is $2M$ so it increases with the
enhancement of $M$. At $M=\triangle /2$, the plateau width reaches
its widest value $\triangle $. By further increasing $M$, the
plateau width becomes narrow and it is equal to $2(\triangle -M)$
when $\triangle /2<M<\triangle $. At $M=\triangle $, the 100\% MR
plateau disappears. When $M>\triangle $ a valley instead of the
plateau appears in the vicinity of $E_{F}=0$. Therefore for a larger $%
\triangle $, it is more favorable for the formation of the 100\% MR
plateau. In experiment, the graphene nanoribbon with its width of
several-ten or sub-ten nanometers has been fabricated
\cite{sci3191229}. Taking a $20nm$-width ribbon for an example, its
$\triangle$ is about $0.12eV$, which is usually much larger than
$M$. In addition, when $E_{F}$ deviates far from $0$ (i.e. the
region of the 100\% plateau), the MR may be quite small (see
figure~\ref{fig5}). This implies that the capability of the
graphene-base spin valve is optimal when it works near the Dirac
point.

\begin{figure}[tbp]
\centering
\includegraphics[bb=75 31 475 413,width=8.4cm]{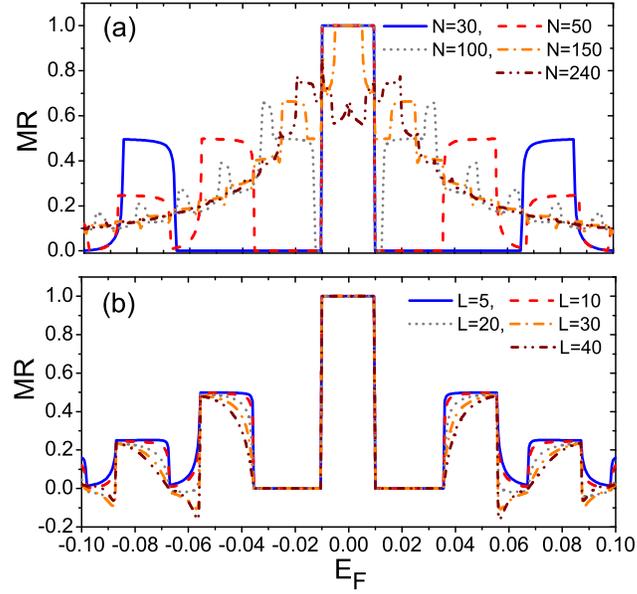}\\%
\caption{(Color online) \textbf{(a)} The MR vs. $E_{F}$ for different width $%
N$ with the length $L=10$ and \textbf{(b)} MR vs. $E_{F}$ for
different length $L$ with the width $N=50$. The magnetization
$M=0.01$. } \label{fig6}
\end{figure}

We study the influence of the size of the ribbon between two FM
leads on the MR. Figure~\ref{fig6}(a) (Figure~\ref{fig6}(b)) shows
the MR as a function of $E_{F}$ for different width $N$ (length $L$)
and a fixed $L$ ($N$). With the increasing of width $N$, the
separation $\triangle $ of the transverse subbands decreases
monotonously (with $\triangle \approx 3\pi t/4N $)
\cite{prb77115408}. In the beginning, the 100\% MR plateau is not
affected until $\triangle /2<M$ (see the curve with $N=30$, $50$,
and $ 100$ in figure~\ref{fig6}(a)). Then with further increasing
$N$, the 100\% MR plateau becomes narrow (see the curve with $N=150$
in figure~\ref{fig6}(a)), and it disappears at the width $N$ with
its $\triangle =M$. At last, when the width of ribbon is very wide
with its $\triangle \ll M$, the MR is very small regardless of other
parameters. This result is the same with recent work studying on the
infinite wide graphene spin valve device \cite{prb76205435}. So it
is favorable to select the narrower nanoribbon to fabricate the spin
valve. On the other hand, the length change of the central region
does not affect the 100\% MR plateau at all (see
figure~\ref{fig6}(b)). Only the shapes of some subplateaus are
slightly changed.

\begin{figure}[tbp]
\centering
\includegraphics[bb=53 110 664 483,width=8.4cm]{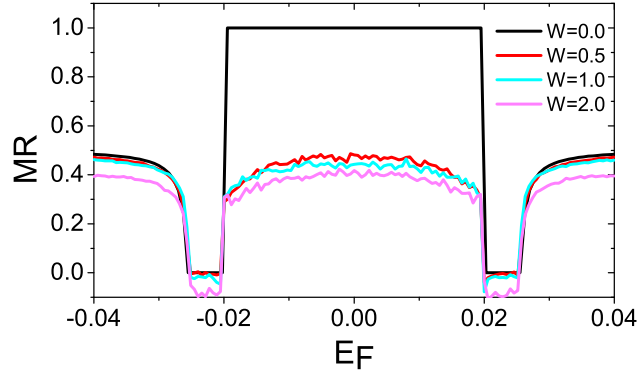}\\%
\caption{(Color online) The MR vs. $E_{F}$ for different disorder strength $%
W $. The parameters are $M=0.02$, $N=50$ and $L=10$. } \label{fig7}
\end{figure}

In the above discussion, we assumed that the Dirac-point energy
$\epsilon _{C}$ of the central region is equal to $\epsilon _{\alpha
}$ of the left and right leads. If $\epsilon _{C}$ departs from
$\epsilon _{\alpha }$, the
conductance $G$ in the antiparallel configuration still is zero when $%
E_{F}\in (-M,M)$ and $0<M<\triangle /2$, because the parity of transverse
wave functions in the left and right FM leads can not match with each other.
However, the conductance $G$ for the parallel configuration is very large ($%
>e^{2}/h$). So the 100\% MR plateau is not affected by $\epsilon
_{C}$ slightly departing from $\epsilon _{\alpha }$.

How is the 100\% MR plateau affected by the disorder? Here, we
consider the Anderson disorder which exists only in the central
graphene region. The on-site energy $\epsilon _{C}$ in the central
region Hamiltonian becomes $\epsilon _{C}+w_{i}$, where $w_{i}$ is
uniformly distributed in the range of $[-W/2,W/2]$.
Figure~\ref{fig7} shows the MR vs. $E_F$ at different disorder
strength $W$, in which every MR curve for $W\not=0$ is averaged over
up to 1000 random configurations. The MR is reduced by the disorder,
but the plateau shape is still hold and the value is about 40\%-50\%
even in quite strong disorder strength.

\begin{figure}[htbp]
\centering
\includegraphics[bb=25 10 725 520,width=8.5cm]{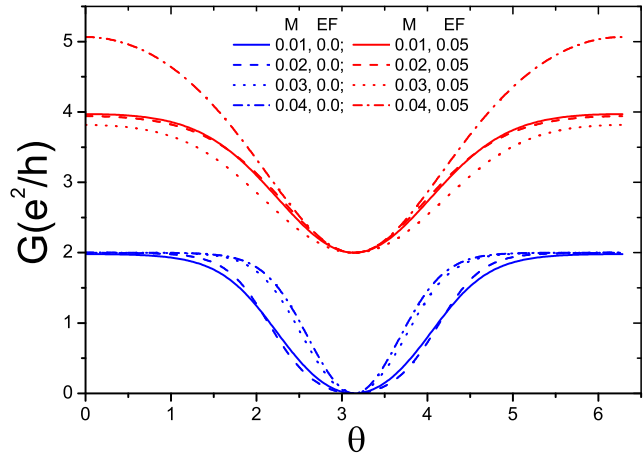}\\%
\caption{(Color online) The conductance $G$ vs. the angle
$\protect\theta $ for different magnetization $M$ and different
Fermi energy $E_{F}$. The size of the central region is $N=50$ and
$L=10$.} \label{fig8}
\end{figure}

\subsection{The conductance for the arbitrary angle $\protect\theta $ of the
left and right magnetizations}

Now we analyze $\theta $ dependence of the conductance. In
figure~\ref{fig8}, the conductance $G$ as a function of the angle
$\theta $ are plotted with combination of $M$ and $E_F$. Here the
shape of the curves $G$-$\theta $ is approximatively a cosine
function regardless of the parameters $M$ and $E_{F}$. At $E_{F}=0$
(i.e. at the center of the $100\%$ MR plateau), the change quantity
of the conductance $G$ for the angle $\theta $ from $0$ to $\pi $ is
about $2e^{2}/h $, which is quite large. About at $\theta =2$, the
derivative $dG/d\theta $
reaches the extreme value, but at the region of $0<\theta <1$, the derivative $%
dG/d\theta $ is quite small. On the other hand, when $E_{F}$
deviates away from the $100\%$ MR plateau, the variation of the
conductance $G$ is also about $2e^{2}/h$ or even larger for some
values of $E_{F}$ (see the curves for $E_{F}=0.05$ in Fig.8),
although the MR is about $50\%$ there. However,
at the region of $MR=0$, the conductance $G$ is independent of the angle $%
\theta $ (not shown here).

\begin{figure}[htbp]
\centering
\includegraphics[bb=189 317 435 460,width=8.5cm]{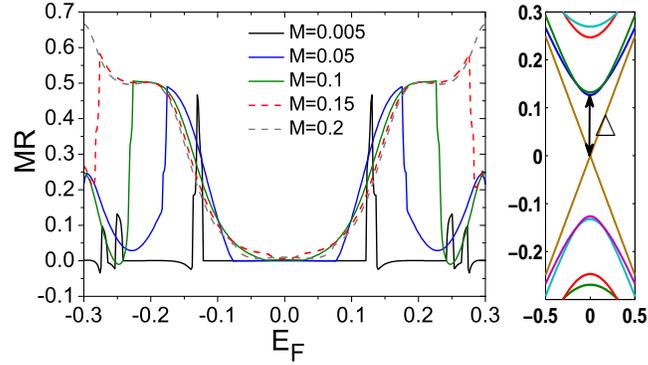}\\%
\caption{(Color online) \textbf{Left panel}: The MR as a function of
$E_{F}$ in the case of the armchair edge with the width $N=41$ and
the length $L=40$ atomic layers in the central region. \textbf{Right
panel}: The energy band structure of an ideal armchair edge graphene
ribbon with the width $N=41$. } \label{fig9}
\end{figure}

\section{The case of armchair edge}

\label{IV}

In this section, we investigate the MR in the armchair edge FM/normal/FM
junction. In the case of the armchair edge, there are two types of energy
band structure \cite{nanores1361}. If the number of transverse atomic layers $%
N$ is equal to $3m$ or $3m+1$ ($m$ is an integer), an energy gap
appears. This gap can be quite large for a narrow graphene
nanoribbon. When the Fermi energy $E_{F}$ is in the gap, the
conductance is always very small regardless of the parallel or
antiparallel configurations. So it is not interesting even if its MR
is very large. For $N=3m+2$, the graphene nano-ribbon is metallic
and its conductance is large.

Figure~\ref{fig9} shows the MR as a function of Fermi energy $E_{F}$
for the metallic armchair edge FM/graphene/FM junction with
$\triangle \approx 0.13t$ for $N=41$. The conductance approaches
$2e^{2}/h$ for both the parallel and antiparallel configurations at
a large range of $E_{F}$ around zero energy.
As a result the MR is very small regardless of $M>\triangle $ and $%
M<\triangle $. On the other hand, when $E_{F}$ is far away from 0 (e.g. $%
E_{F}>\triangle $), the MR can be over 50\% in some regions of
$E_{F}$. Anyway, the property of MR of the zigzag edge ribbon is
much better than that of the armchair edge ribbon. So it is more
favorable to fabricate the spin valve devices by using the zigzag
edge ribbon. In experiment, the graphene nanoribbon with the
specific parity edge has already been successfully fabricated
\cite{sci3191229}.

\section{Conclusions}

\label{V}

In summary, we have studied the electronic transport and
magnetoresistance (MR) in graphene-based
ferromagnetic/normal/ferromagnetic junction where the finite width
and the arbitrarily relative orientation between the lead
magnetizations are taken into account. For the zigzag edge case, the
conductance for the parallel configuration is always larger than
$e^{2}/h$ under any parameter condition, but for the antiparallel
configuration the conductance is exactly zero when Fermi energy is
around the Dirac point due to the band-selective rule. This leads to
a 100\% MR plateau. The 100\% MR plateau is almost not affected by
the length of the central graphene region. With the increasing of
the width of the ribbon, the 100\% MR plateau is kept stable in the
beginning, but is smeared at the wide ribbon limit. In the presence
of the disorder, the MR is slightly suppressed, but can still keep
the plateau shape with the value about $50\%$ even in quite large
disorder strengths. When the orientation between magnetizations of
two ferromagnetic leads is arbitrary, the conductance versus the
relative angle $\theta $ is similar to a cosine formation. What's
more, for the armchair edge case, the MR is relatively small.
Therefore, it is more favorable to fabricate the graphene-based spin
valve devices by using the zigzag edge graphene ribbons.

\section*{Acknowledgments}

We gratefully acknowledge the financial support from NSF-China under
Grants Nos. 10525418, 10734110, and 10821403, the 973 Program
Project No. 2009CB929103, and the Research Grant Council of Hong
Kong under Grant No. HKU 7042/06P and HKU 10/CRF/08. Shu-guang Cheng
was supported by the Science Foundation of Northwest University
(No.09NW29).

\end{document}